# Impact of post deposition annealing in $O_2$ ambient on structural properties of nanocrystalline hafnium oxide thin film

Shilpi Pandey[*,1,2], Prateek Kothari[1], Sunil Kumar Sharma[3], Seema Verma[2], K.J. Rangra[1]
[1] CSIR-Central Electronics Engineering Research Institute, Pilani, India
[2] Banasthali Vidyapith, Banasthali, India
[3] Solid State Physics Laboratory, New Delhi, India

**Abstract**

**In the present work, $HfO_2$ thin film (100nm) has been deposited by sputtering technique and annealed at various temperatures ranging from 400 to 1000 ºC (in step of 200 ºC) in $O_2$ ambient for 10 min. The samples have been characterized using XRD, FTIR, EDAX, AFM and Laser Ellipsometer. The impact of annealing temperatures in $O_2$ ambient on structural properties such as crystallite size, phase, orientation, stress have been studied using XRD. The Hf-O phonon peaks in the infrared absorption spectrum are detected at 512, 412 $cm^{-1}$. The stretching vibration modes at 720 $cm^{-1}$ and 748 $cm^{-1}$ correspond to $HfO_2$. AFM data show mean grain size in the range of 38 nm – 67 nm. The film reveals variation in structural properties, which appears to be responsible for variation in oxygen percentage, refractive index (1.96-2.01) at 632 nm wavelength and roughness (6.13nm-16.40nm). Annealing temperature as well as ambient condition has significant effects on stress, crystal size and thus the arrangement of atoms. For good quality film, annealing temperature larger than 600 ºC is desired.**

*Keywords:* **Hafnium Oxide, Annealing, X-ray diffraction, AFM, FTIR, Ellipsometer, Crystallite size, Strain, Grain size**

## 1. Introduction

Hafnium oxide has been reported as potential contender among several other high-k dielectrics such as barium strontium titanate, zirconium titanate etc. The primary factors responsible for this are its outstanding electrical properties like wide band gap, high refractive index, high dielectric constant and better chemical stability i.e. excellent process compatibility with concurrent IC technology [1-2].

The miniaturization of electronic devices viz., metal-insulator-semiconductor, RF (radio frequency) micro-electro-mechanical systems (MEMS) capacitive switch, RF MEMS phase shifter, dynamic random access memory, electro luminescent devices etc., along with improved performance ( high speed, reduce size and low power consumption ) have been demonstrated by replacing low-k dielectrics such as $SiO_2$, $Si_3N_4$ or its oxynitrides by high-k dielectrics [1, 3, 4].

Employing hafnium oxide in RF MEMS capacitive switch drastically improves RF performance in terms of large down state capacitance which results in better isolation, high capacitance ratio as well as reduction in size. The study conducted by Yi Zhang et al. [5], demonstrated RF MEMS capacitive switch using hafnium oxide with isolation -40 dB in the frequency range 4-35 GHz. Hafnium oxide based RF MEMS capacitive switch with better isolation -60 dB at 35 GHz as well as capacitance ratio 43 was exhibited by X.J. He et al. [8]. Various research groups have reported characterization of hafnium oxide [5-7]. However, the detailed information of the microstructure and morphology of $HfO_2$ film have not been studied well for RF MEMS devices. Therefore, it is important to look into the structure of film.

*Corresponding Author. Contact: +91-9530478542
*E-mail Address:* er.shilpi@gmail.com , shilpi@ceeri.ernet.in (Shilpi Pandey)

Many deposition techniques such as chemical vapor deposition [9], ion-beam evaporation [10], RF sputtering [11-12], pulsed laser deposition [13] for $HfO_2$ thin film have been reported. In the present work, RF sputtering has been employed due to its low temperature processing, high deposition rate, good step coverage [1, 14-15]. Further the as-deposited samples have been further annealed at 400 °C, 600 °C, 800 °C and 1000 °C in $O_2$ ambient for 10 min. The deposition technique and post deposition annealing have significant impact on structural properties [4]. Hafnium Oxide exhibits three different phases: monoclinic, tetragonal and cubic, depending upon process parameters, though the most stable phase is monoclinic [2]. To obtain thin film with high dielectric constant, it is necessary to have a correct combination of various structural properties e.g. phase, texture and stress [16-18]. This paper investigates, the effect of post deposition annealing in $O_2$ ambient, on structural and morphological properties.

## 2. Experimental Details

A 2 inch low resistive p-type Si (100) substrate has been taken for the process. After standard cleaning treatment, substrate has been subjected to moisture bake at 120 °C to prepare it for subsequent $HfO_2$ thin film deposition using MRC 8620J sputter system. Initially, vacuum chamber has been evacuated to base pressure 3e-6 Torr. $HfO_2$ target of 2 inch diameter and 99.95 % pure has been employed for sputtering process. The target has been kept at 8 cm distance from the substrate. Before deposition, $HfO_2$ target has been pre-sputtered for 10 mins using Ar alone with shutter above the gun closed. The deposition has been carried out for 35 mins with sputtering power 250 W to achieve 100 nm thin $HfO_2$. The chamber pressure has been maintained at 3 mTorr during sputtering. Further the samples have separately been annealed in quartz tube furnace at temperatures 400 °C, 600 °C, 800 °C, 1000 °C respectively for 10 minutes each in $O_2$ ambient.

The structural measurements have been characterized by X-ray diffraction using Bruker D8 Advance X-ray diffractometer system. The incident beam optics consists of a Cu Kα radiation source (λ=1.5406Å). The crystallite size of 100 nm $HfO_2$ thin film has been calculated using well known Scherrer's Eq.2.1 [19-20].

$$D = \frac{k\lambda}{\beta \cos \theta} \qquad 2.1$$

where $D$ is the crystallite size, $k$(=0.9) is the crystal constant, $\lambda$ is the wavelength of X-ray used, $\beta$ is the broadening of diffraction line measured at half of its maximum intensity and $\theta$ is the angle of diffraction [19-20]. Bragg's law has been used to calculate the interplanar spacing, $d_{(hkl)}$, from $2\theta_{(hkl)}$ as shown in Eq. 2.2.

$$d = \frac{\lambda}{2 \sin \theta} \qquad 2.2$$

A Bruker Tensor 37 type Fourier transform infrared (FTIR) spectrometer has been used to obtain bond information of the $HfO_2$ thin films. The samples have been studied in the range of 1200-400 $cm^{-1}$ by FTIR spectroscopy.

The Elemental or Energy dispersive X-ray spectroscopy (EDAX) has been used to detect elements present in significant quantity (quantitative determination of bulk element composition). The EDAX analysis of $HfO_2$ film deposited on silicon substrate, has been carried out on JEOL SEM system operated at 16 kV accelerating voltage. Surface morphology of $HfO_2$ thin film has been studied by Nova Atomic Force Microscope (AFM). The refractive index and thickness have been measured by laser ellipsometer SENTECH SE500 using laser radiation of 632 nm wavelength.

## 3. Results and Discussion

### 3.1. Crystallographic analysis

The crystal structure and orientation of the $HfO_2$ samples have been studied using X-ray diffraction (XRD) patterns. Fig. 1(a) shows the typical XRD patterns of as-deposited and annealed $HfO_2$ thin films at 400 ºC, 600 ºC, 800 ºC and 1000 ºC in $O_2$ ambient which specify that, the $HfO_2$ is purely crystalline in nature. The XRD pattern of as-deposited $HfO_2$, contains peak at Bragg's angle 2θ=28.45895, assigned to (-111) crystallographic plane which indicates the presence of small nano crystallites [33]. Minor peaks of other orientation are also present due to monoclinic crystallites. Therefore, anisotropy exists and (-111) crystallographic plane exhibits lowest strain energy [2, 21].

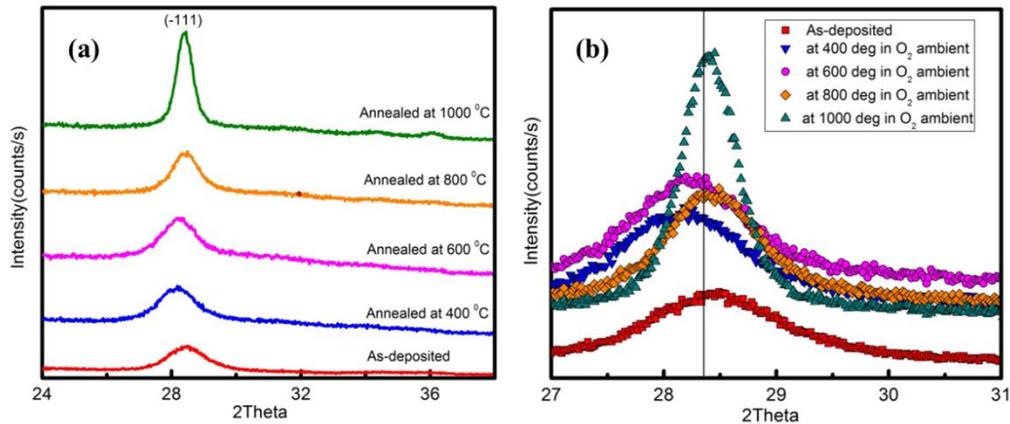

**Fig. 1.** (a) XRD patterns of $HfO_2$ films as-deposited and annealed at various temperatures in $O_2$ ambient. (b) High resolution XRD scans of monolithic (-111) peaks of $HfO_2$ films.

Crystallite size and preferred orientation along (-111) planes increase with increase in annealing temperatures. XRD pattern of annealed $HfO_2$ at 1000 ºC indicates highly oriented nature. In the present case, increasing temperature favors the preferred orientation along (-111) while minimizing the strain energy. For further analysis of growth process, crystallite size and lattice mismatch at nanoscale dimension, high resolution scans have been performed on (-111) plane as shown in Fig. 1(b). Due to the shift and broadening of diffraction peaks, significant change occurs in crystallite size and strain [21]. The diffraction peak shifts to higher angle (2θ) in as-deposit $HfO_2$ thin film compare to standard position at 28.347 [22]. This indicates to the fact that the contraction of the lattice occurs with 0.38 % compressive strain. At lower annealing temperature i.e. 400 ºC and 600 ºC, 2θ shifts to lower side of standard value which points to the expansion of the lattice with tensile strain of 0.58 % and 0.39 %, respectively. With further increase in annealing temperature at 800 ºC and 1000 ºC, peak again shifts to higher 2θ which attribute to lattice contraction with compressive strain 0.30 % and 0.19 %, respectively.

Most of the existing research work have studied the effect of stress on crystal arrangements and have significantly discussed only lattice expansion after annealing [21]. None of the prior works have observed both trends i.e. lattice expansion and lattice contraction at annealing temperature. However, we have observed both the two different trends i.e. lattice expansion at lower annealing temperature and lattice contraction at higher annealing temperature. It is anticipated that the lattice expansion has occurred due to dominant repulsive force between $Hf^{4+}$-$Hf^{4+}$ atoms. This signifies that there is deficiency of oxygen. However at high annealing temperature, lattice contraction exists due to strong attraction force between $Hf^{4+}$-$O^{2-}$ dipoles which in effect points to oxygen efficiency. The same is exhibited in results from FTIR and laser ellipsometry analysis which is discussed in further section. The crystallite size obtained for (-111) crystallographic plane of as-deposited $HfO_2$ thin film is ≈ 6.12nm, which matches very well with previously reported article [21]. With increase of the annealing temperature, the crystallite size increases, nevertheless the *d* spacing decreases as shown in Fig. 2(a) & Table 1.

**Table 1**
Crystallographic properties of 100nm $HfO_2$ thin film for most intense (-111) peak

| Annealing Temperature | FWHM | 2θ(deg) | *D*(nm) | *d*(nm) |
|---|---|---|---|---|
| as-deposited | 1.33837 | 28.45895 | 6.12 | 0.3132 |
| 400 ºC | 1.272 | 28.17662 | 6.4 | 0.3163 |
| 600 ºC | 1.16311 | 28.23311 | 7.04 | 0.3157 |
| 800 ºC | 0.94207 | 28.43506 | 8.69 | 0.3135 |
| 1000 ºC | 0.62054 | 28.40323 | 13.2 | 0.3138 |

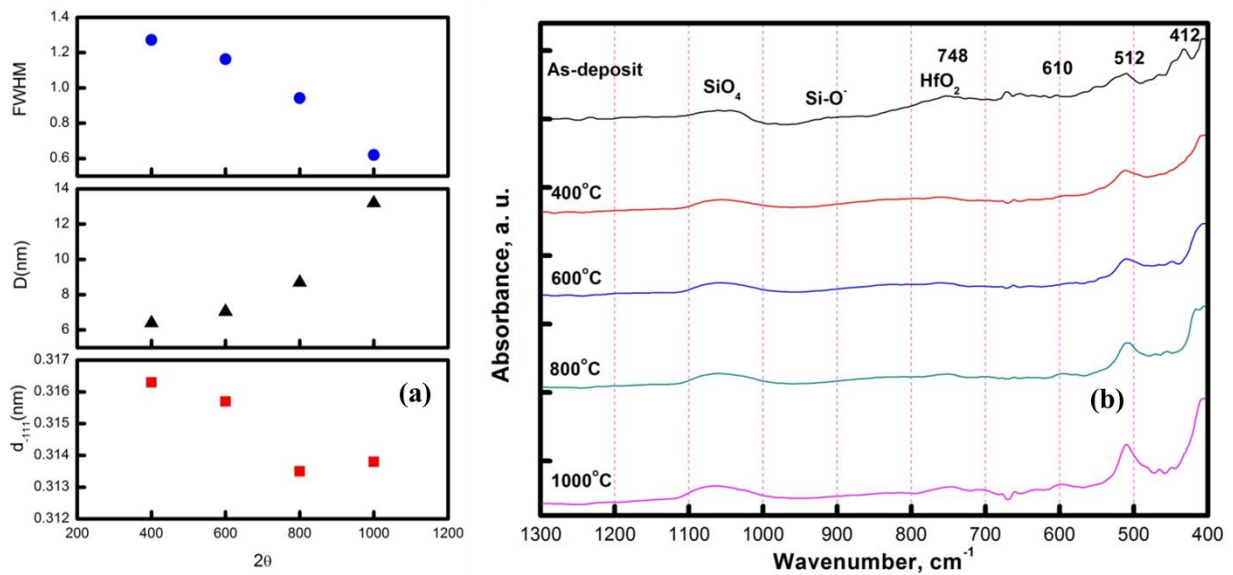

**Fig. 2.** (a) Variation of FWHM, D (crystallite size) and *d*(-111) spacing with annealing temperature. (b) Infrared absorption spectra for HfO$_2$ thin film as-deposited and annealed at various temperatures in O$_2$ ambient.

### 3.2. FTIR spectra of sputter deposited HfO$_2$ thin film

Fig. 2(b) shows Fourier Transform Infrared (FTIR) spectrum of HfO$_2$ thin film. The bonding structures of HfO$_2$ films have been identified in the 1200-400 cm$^{-1}$ spectral region. FTIR spectrum of hafnium oxide film shows broad absorption band between 1100 and 1000 cm$^{-1}$ which corresponds to transverse optical component of asymmetrical stretch of SiO$_4$[3, 23-24].

According to report Neumayer et al. [25], a wide absorption band between 1200 and 810 cm$^{-1}$ indicated that it was due to three components viz., asymmetric stretch of SiO$_4$ between 1180 and 1080 cm$^{-1}$ [26], absorption peak at ≈970 cm$^{-1}$ was attributed to HfSiO, and at ≈880 cm$^{-1}$ was assigned to Si-O$^-$. Absorption peak, indicated at ≈1105 cm$^{-1}$ was assigned to interstitial oxygen in the Si bulk [3, 27].

However in this research work, peaks at ≈1105 cm$^{-1}$, 970 cm$^{-1}$ and 880 cm$^{-1}$ have not been found. T.C. Chen et al. reported the presence of a peak lying between 934 cm$^{-1}$ and 838 cm$^{-1}$ corresponding to Si-O$^-$ [28]. However in as-deposited film, a peak has been found at 921 cm$^{-1}$, later which has disappeared after annealing treatment. It is apparent that when the annealing temperature increases, strength of absorption peak assigned to Si-O vibration becomes weak. The weak peak detected at 610 cm$^{-1}$ for the film annealed at 800 and 1000 °C is related to absorption of a Si phonon [24]. Also, the wide peak lying at 748 cm$^{-1}$ corresponds to HfO$_2$ [3, 24]. The other main peaks around 512, 412 cm$^{-1}$ are due to Hf-O chemical bonds [3, 23, 29-31]. With increase in annealing temperature, more oxygen is absorbed by thin film and therefore Hf-O bond peaks increase due to oxidation of HfO$_2$ thin film. At 412, 512 cm$^{-1}$, absorption of photon increases with increase in annealing temperature which reveals strengthening of Hf-O bonds as shown in Fig 2 (b). Using XRD analysis, the same dominant attraction force between Hf$^{4+}$- O$^{2-}$ dipoles (which signifies more oxygen is absorbed by thin film) is observed at 800 °C and 1000 °C annealing temperature.

### 3.3. Elemental composition analysis of HfO$_2$ film

EDAX spectrum (shown in Fig. 3) represents the different elements present in the thin film. The data in Fig. 3 is shown with no smoothing, filtering or processing of any kind. The EDAX spectrum shows clear peaks corresponding to the 72 Hf L (7.89 keV) line, 72 Hf M line (1.64 keV) and 8 O K line (0.52 keV). The 14 Si K line (1.74 keV) peak is observed in the EDAX spectrum is due to silicon substrate. No other peak is observed over the entire 0 keV to 20 keV detection window.

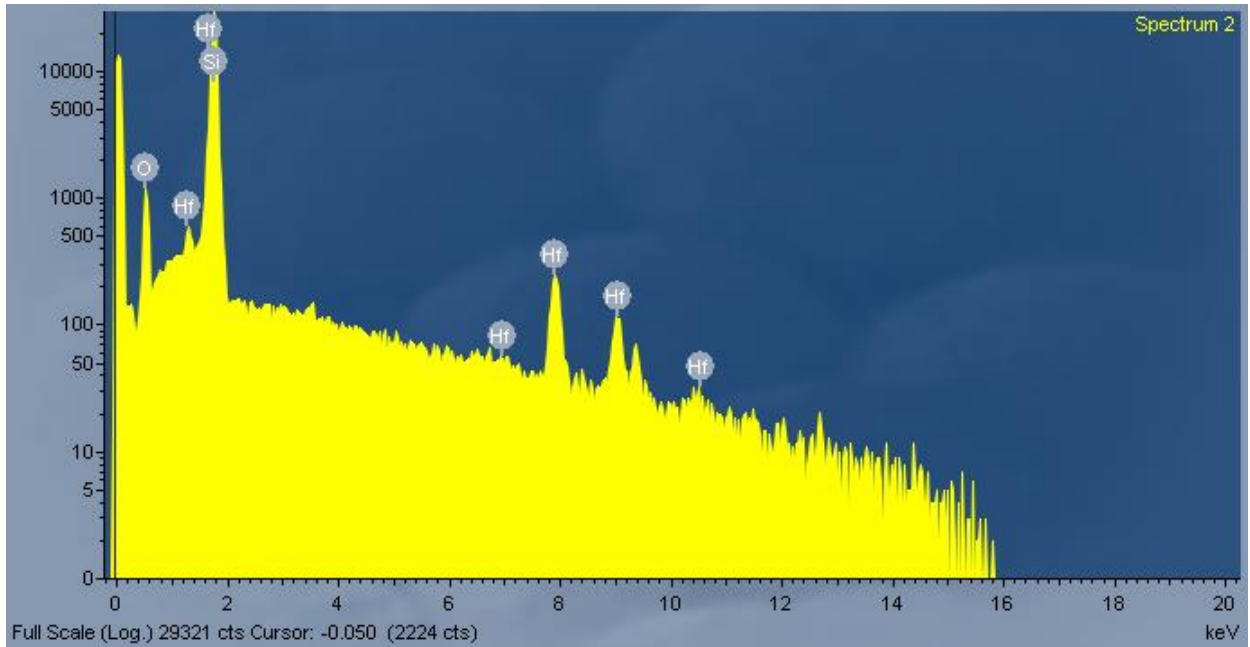
**Fig. 3.** Chemical composition of $HfO_2$ is determined by EDAX.

### 3.4. Surface Morphology

The surface morphology of as-deposited and annealed $HfO_2$ thin films have been analyzed using atomic force microscopy in tapping mode. The films are scanned over 1μm x 1μm at 3.656V, 1 Hz frequency. Fig. 4 shows 2-D and 3-D AFM images of as-deposited and annealed $HfO_2$ thin films. AFM images have been obtained at different locations of wafer which show that the film is homogeneous, free of cracks and pinholes. The root mean square (RMS) roughness, average roughness and mean grain size are shown in Table 2. Annealing temperature, ambient conditions and stress have significant impact on grain size as well as surface roughness [32]. Most of the research groups have studied a normal trend of increasing grain size after annealing [34]. Instead, we have analyzed different trends of grain size. AFM data shows that the film's RMS roughness reduces maximum at 600 ºC annealing temperature. The reduction in the size of nanoparticles is observed at 600 ºC annealing temperature. However, with further increase in annealing temperature from 600 ºC to 800 ºC, the smaller nanoparticles start combining together in order to form a larger nanoparticle. It is clear that rearrangement of nanoparticles at annealing temperature is due to presence of stress in the film.

**Table 2**
Roughness and grain parameters of $HfO_2$ thin films

| $HfO_2$ thin film | RMS roughness (nm) | Average roughness (nm) | Mean grain size (nm) |
|---|---|---|---|
| as-deposited | 7.33 | 6.13 | 37.856 |
| 400 ºC | 19.70 | 17.20 | 53.202 |
| 600 ºC | 8.46 | 7.09 | 48.359 |
| 800 ºC | 14.80 | 12.40 | 66.000 |
| 1000 ºC | 13.2 | 16.4 | 67.364 |

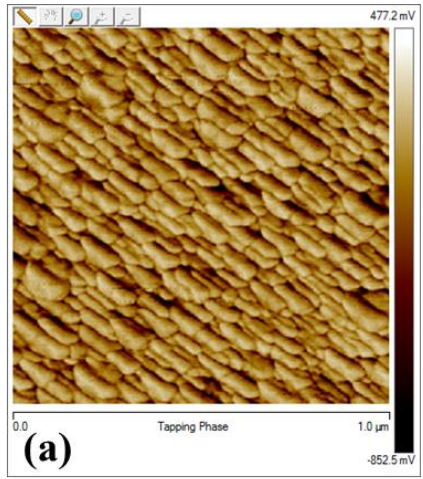 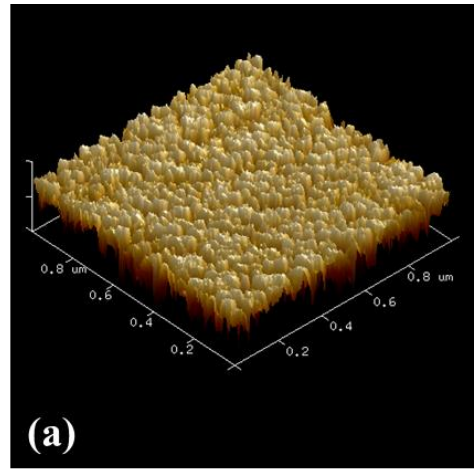

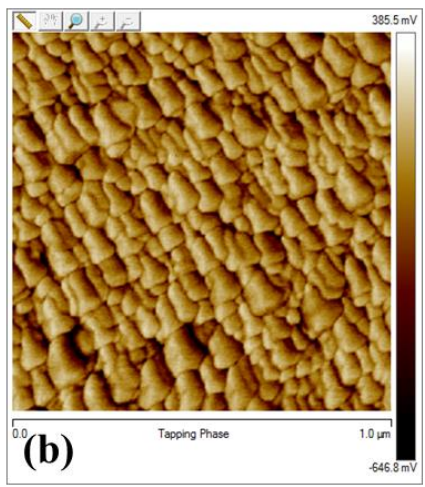 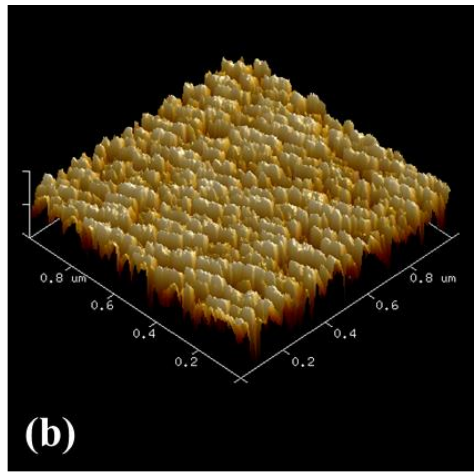

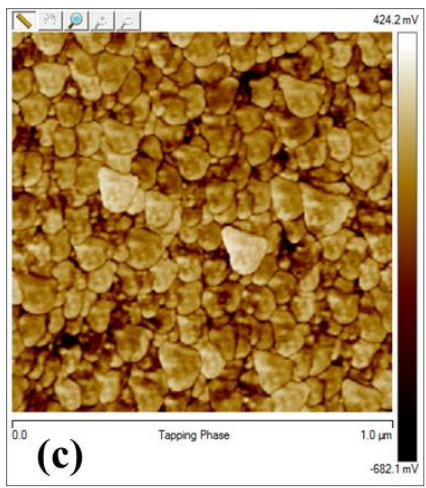 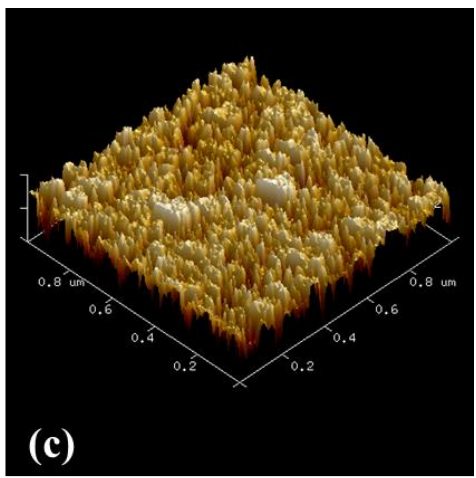

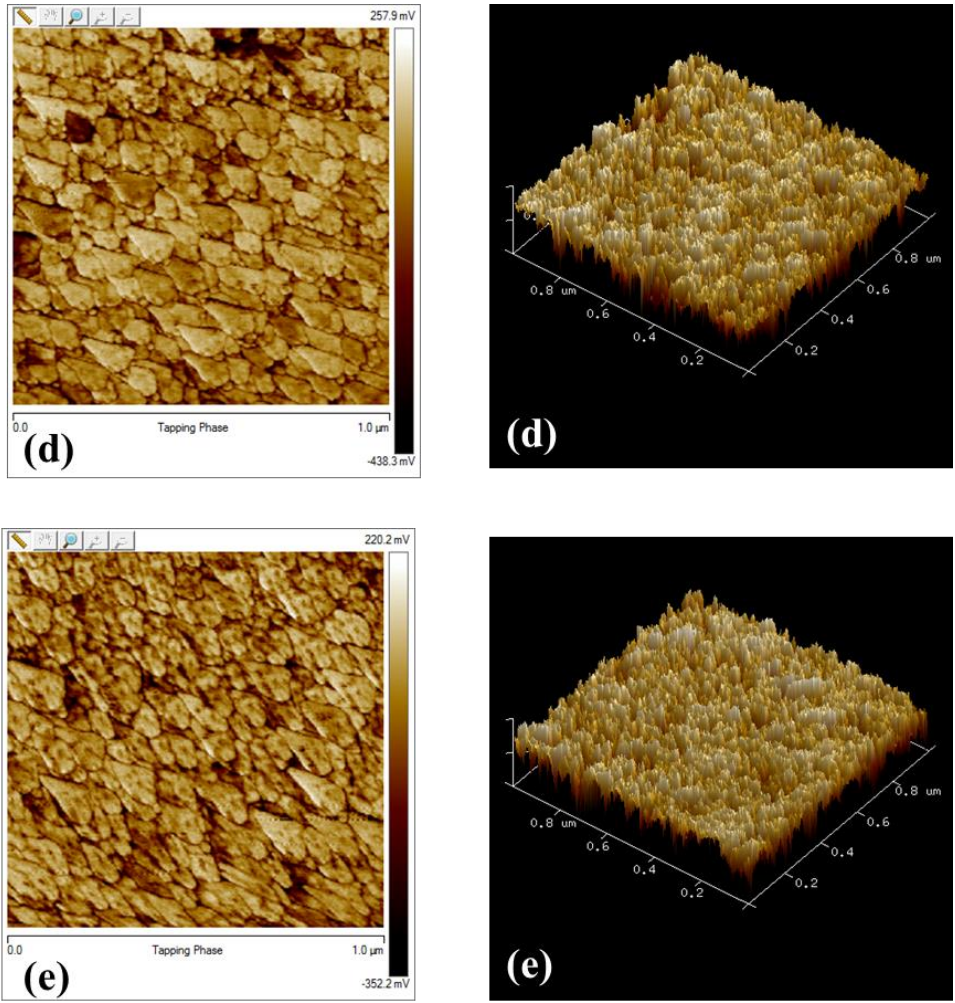

**Fig. 4.** 2-D and 3-D AFM images of HfO$_2$ thin films: (a) as-deposited (b) 400 °C (c) 600 °C (d) 800 °C (e) 1000 °C annealed in O$_2$ ambient.

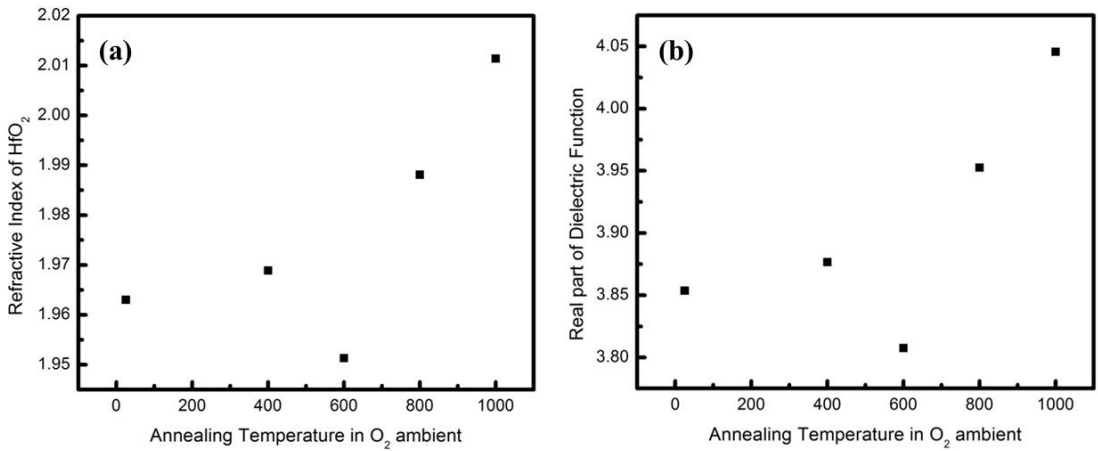

**Fig. 5.** (a) Dependence of refractive index (n) on annealing temperature in O$_2$ ambient, at wavelength 632 nm. (b) Real part of dielectric constant, ε, at wavelength 632 nm on annealing temperature in O$_2$ ambient.

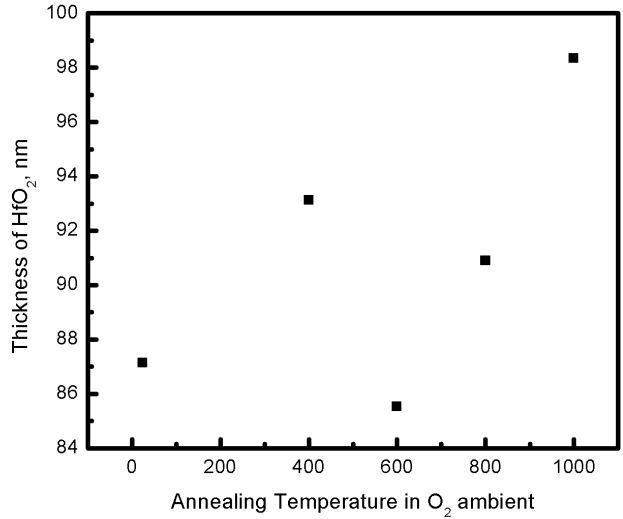

Fig. 6. Thickness of $HfO_2$ thin film versus annealing temperature, at wavelength 632 nm.

### 3.5. Laser Ellipsometer

Laser Ellipsometer is a high performance tool to measure refractive index and thickness of materials. It is widely known that the refractive index is closely related to the structural properties and optical density of the film [1]. The thickness of $HfO_2$ thin film is shown in Fig. 6. Fig. 5 shows refractive index and real part of dielectric constant of $HfO_2$ thin film as-deposited and annealed at various temperatures in $O_2$ ambient.

It is observed that refractive index decreases initially at 600 °C annealing temperature which signifies that the film is less optically dense. This could have occurred due to stress (i.e. lattice expansion), which implies dominant repulsive force between $Hf^{4+}$-$Hf^{4+}$ atoms and therefore reduction in oxygen atoms. Also reduction in the size of nanoparticles at 600 °C annealing temperature is confirmed from AFM data. At 800 °C and 1000 °C annealing temperature, refractive index increase which indicates that the film is denser (optically). It is considered to be caused by lattice contraction, resulting in increase in attraction force between $Hf^{4+}$-$O^{2-}$ dipoles. The same is concluded from XRD and FTIR analysis. The variation in grain size and thickness of $HfO_2$ thin film with increase in annealing temperature is confirmed from the AFM data and thickness plot (shown in Table 2 and Fig. 6). The annealing temperature as well as ambient has significant impact on thickness of thin film. At low annealing temperature, the variation in film thickness is due to the rearrangement of grains and then densification of thin film occurs which is also confirmed using AFM analysis. At high annealing temperature, film thickness increases due to oxygen diffusion in $HfO_2$ thin film. Thus oxidation of $HfO_2$ thin film has been occurred and at 1000 °C film is fully oxidized [35]. It is observed that the nature of polarization is different due to presence of stress in crystalline $HfO_2$ thin film [32]. Since the sputtered thin film is transparent at 632 nm, therefore, dielectric constant's imaginary part goes to zero (no extinction coefficient) and the dielectric constant becomes equal to the square of refractive index ($\varepsilon_1=n^2$) [3]. Fig. 5(b) shows the dielectric constant's real part ($\varepsilon_1$) of as-deposited $HfO_2$ and annealed $HfO_2$.

## 4. Conclusion

The current study reveals that HfO$_2$ thin films are poly crystalline in nature and exhibit monoclinic structure. These monoclinic structures are highly oriented along (-111) direction. The analysis from XRD indicates that crystallite size increases from 6.12 nm to 13.2 nm with increase in annealing temperature. It is inferred that structural properties viz. stress, crystallite size, grain size of thin film depend strongly on annealing temperature. Mean grain size of each of the individual as-deposited and annealed films measured using AFM are found to be in the range of 37.856 nm - 67.364 nm. The experimental results conclusively suggest that annealing temperature needs to be larger than 600 ºC for a good quality HfO$_2$ thin film with high refractive index, large crystallite size, grain size.

## 5. Acknowledgement

I am grateful to Dr. Chandershekhar, Ex-Director, CSIR- CEERI, Pilani for giving me the opportunity to carry out the research work at CSIR-CEERI Pilani. I sincerely thank Mr. Triloki, Senior Researcher, BHU, Varansai for his invaluable guidance and help. I also express my thanks to Mr. Ashok, Technical Assistant, Banasthali University, Banasthali and Mr. Sanjeev Kumar, Scientist, CSIR-CEERI Pilani, for carrying out XRD and AFM experiment, respectively. The authors acknowledge the financial assistance under network project-PSC0201, Council of Scientific and Industrial Research (CSIR). Author is grateful to HRDG, CSIR, for granting Senior Research Fellowship.